\documentclass[prl,twocolumn,preprintnumbers,amsmath,amssymb]{revtex4}
\usepackage{graphicx}

\begin{document}

\title{  Angle resolved photon spectrum and quasiparticle excitation spectrum in an exciton superfluid  }
\author{ Jinwu Ye$^{1}$, T. Shi$^{2}$ and  Longhua Jiang $^{1}$   }
\affiliation{$^{1}$Department of Physics, The Pennsylvania State
University, University Park, PA, 16802, USA  \\
 $^{2}$Institute of
Theoretical Physics, Chinese Academy of Sciences, Beijing, 100080,
China }
\date{\today }

NSF-KITP 09-42

\begin{abstract}
 There have been extensive
experimental search for possible exciton superfluid  in
semiconductor electron-hole bilayer systems below liquid Helium
temperature. However, exciton superfluid are meta-stable and will
eventually decay through emitting photons. Here we show that the
light emitted from the excitonic superfluid has unique and unusual
features not shared by any other atomic or condensed matter systems.
We evaluate angle resolved photon spectrum, momentum distribution
curve, energy distribution curve and quasiparticle excitation
spectrum in the exciton superfluid and comment on relevant
experimental data in both exciton and exciton-polariton systems.
\end{abstract}

\maketitle

{\bf 1. Introduction.} An exciton is a bound state of  an electron
and a hole. Exciton condensate was first proposed more than 3
decades ago as a possible ordered state in solids
\cite{blatt,kel,loz}. But so far, no exciton superfluid phase has
been observed in any bulk semi-conductors. Recently, degenerate
exciton systems have been produced by different experimental groups
with different methods in quasi-two-dimensional
semiconductor $GaAs/AlGaAs $ coupled quantum wells structure \cite%
{butov,snoke,field1,field2,bell}. When the distance between the two
quantum wells is sufficiently small, an electron in the conduction
band in one quantum well and a hole in the valence band in the other
quantum well could pair to form an indirect exciton which behaves as
a boson in dilute exciton density limit.
 It was established that the indirect excitons in EHBL has at
least the following advantages over the excitons in the bulk: (1)
Due to the space separation of electrons and holes, the lifetime
$\tau_{ex} $ of the indirect excitons is  $10^{3}\sim 10^{5} $
longer than that of direct ones, now it can be made as long as
microseconds. (2) Due to the relaxation of the momentum conservation
along the $\hat{z} $ direction, the
thermal lattice relaxation time $\tau_{L} $ of the indirect excitons can be made as $%
10^{-3} $ that of bulk excitons, so $\tau_{ex} \gg \tau_{L} $ is
well satisfied. (3) The repulsive dipole-dipole interaction is
crucial to stabilize the excitonic superfluid against the competing
phases such as bi-exciton formation and electron-hole plasma phase.
So EHBL is a very promising system to observe BEC of in-direct
excitons. The quantum degeneracy temperature of a two dimensional
excitonic superfluid (ESF) can be estimated to be $T^{ex}_{d} \sim 3
K $ for
exciton density $n \sim 10^{10} cm^{-2} $ and effective exciton mass $m \sim 0.22 m_{0} $ where $%
m_{0} $ is the bare mass of an electron, so it can be reached easily
by $He $ refrigerator. Indeed, as temperature is decreased from $
\sim 20 K $ to $ \sim 1.7 K $,  the spatially and spectrally
resolved PL peak density centering around the gap $ E_{g} \sim 1.545
eV $ increases \cite{butov}, the exciton cloud size decreases to $
L\sim 30 \mu m $, the peak width shrinking to $ \sim 1 meV $ at the
lowest temperature $ \sim 1.7 K $. All these facts indicate a
possible formation of exciton condensate around $ 1.7K $.


In this paper, we will study quantum nature of photons emitted from
the excitonic superfluid phase in semiconductor electron-hole
bilayer systems. We comment on current PL experimental data and also
propose possible future experiments such as angle resolved power
spectrum to test the existence of exciton condensate in the EHBL
system. The phase sensitive homodyne measurement to detect the two
mode squeezing spectrum and HanburyBrown-Twiss type of experiments
to detect two photon correlations and photon statistics will be
presented in a separate publication \cite{un}. In this paper, for
simplicity, we ignore the spins of excitons, the effects of a trap
and disorders. Their effects are important and will be investigated
in separate publications.

{\bf 2. Exciton-Photon Hamiltonian.  }  The total Hamiltonian in
grand canonical ensemble is the sum of excitonic superfluid part,
photon part and the coupling between the two parts $H_{t}=H-\mu
N_{t}=H_{sf}+H_{ph}+H_{int}$ where :
\begin{eqnarray}
H_{sf} &=&\sum_{\vec{k}}(E_{\vec{k}}^{ex}-\mu )b_{\vec{k}}^{\dagger }b_{\vec{%
k}}+ \frac{1}{2 A} \sum_{\vec{k}\vec{p}\vec{q}}V_{d}(q)b_{\vec{k}-\vec{q}}^{\dagger }b_{%
\vec{p}+\vec{q}}^{\dagger }b_{\vec{p}}b_{\vec{k}}  \nonumber \\
H_{ph} &=&\sum_{k}\omega _{k} a_{k}^{\dagger }a_{k}  \nonumber \\
H_{int} &=&\sum_{k}[ig(k)a_{k}b_{\vec{k}}^{\dagger }+h.c.].
\label{first}
\end{eqnarray}%
 where  $ A $ is the area of the EHBL, the exciton energy $ E_{\vec{k}}^{ex} = \vec{k}^{2}/2M +
 E_{g}-E_{b}$, the photon frequency
 $ \omega_{k}=v_{g}\sqrt{k_{z}^{2}+\vec{k}^{2}}$ where
 $v_{g}=c/\sqrt{\epsilon } $ with $ c $  the light speed in the
 vacuum and $\epsilon \sim 12$  the dielectric constant of $GaAl$,
 $ k=( \vec{k},k_z) $ is the 3 dimensional momentum,
 $V_{d}(\vec{q}) = \frac{ 2 \pi e^{2} }{ \epsilon q }( 1- e^{-qd} ) $ is the dipole-dipole interaction between the
 excitons \cite{ye}, $V_{d}(\left\vert \vec{r}\right\vert \gg
d)=e^{2}d^{2}/\left\vert \vec{r}\right\vert ^{3}$ and $  V_{d}(q=0)
= \frac{ 2 \pi e^{2} d}{ \epsilon } $ leads to a capacitive term for
the density fluctuation \cite{psdw}. The $ g(k) \sim
\vec{\epsilon}_{k\lambda }\cdot \vec{D}_{k} \times L^{-1/2}_{z} $ is
the coupling between the exciton and the photons  where $ L_{z}
\rightarrow \infty $ is the normalization length along the $ z $
direction. Obviously, only the photon polarization in the plane
spanned by $ k $ and $ \vec{D}_{k}$ contributes. Note that the
transition dipole moment $ \vec{D}_{k} $ from the conduction band to
the valence band at a momentum $ k $ is completely different from
the static dipole moment in the dipole-dipole interaction
$V_{d}(\vec{q})$ in Eqn. \ref{first}.

In the dilute limit, $V_{d}$ is relatively weak, so we can apply
standard Bogoloubov approximation to this system. So in the ESF
phase, one can decompose the exciton operator into the condensation
part and the quantum fluctuation part above the condensation $
b_{\vec{k}}=\sqrt{N}\delta _{\vec{k}0}+\tilde{b}_{\vec{k}}$.  Upto
the quadratic terms, the exciton Hamiltonian $H_{sf}$ can be
diagonalized by Bogoliubov transformation:
$H_{sf}=E(0)+\sum_{\vec{k}}E(\vec{k})\beta _{\vec{k}}^{\dagger }\beta _{\vec{k%
}} $ where $E(0)$ is the condensation energy, $
E(\vec{k})=\sqrt{\epsilon _{\vec{k}}[\epsilon _{\vec{k}}+2\bar{n}V_{d}(\vec{k%
})]} $, $ \beta _{\vec{k}}=u_{\vec{k}}\tilde{b}_{\vec{k}}+v_{\vec{k}}\tilde{b}_{-\vec{k%
}}^{\dagger } $ is the Bogoliubov quasi-particle annihilation
operators with  the two coherence factors $ u_{\vec{k}}^{2} ( v_{\vec{k}}^{2} ) =\frac{\epsilon _{\vec{k}}+\bar{n}V_{d}(\vec{k})}{2E(\vec{k%
})} \pm \frac{1}{2} $. The linear term in $\tilde{b}_{k}$ is
eliminated by setting the chemical potential $ \mu =E_{0}^{ex}+
\bar{n} V_{d}(0)= ( E_g-E_b) + \bar{n} V_{d}(0) $. {\sl In a
stationary state, the $ \mu $ is kept fixed at this value
\cite{andrei}}. As $\vec{k}\rightarrow 0$, $E(\vec{k})=u \left\vert
\vec{k} \right\vert $ shown in Fig.1a where the velocity of the
quasi-particle is $u=\sqrt{ \bar{n}V_{d}(0)/M} =   \sqrt{ \frac{ 2
\pi e^{2} d \bar{n} }{ \epsilon M } } $.
Even at $ T=0 $, the number of excitons out of the condensate is: $
n^{\prime}(T=0) = \frac{1}{S} \sum_{\vec{k}}
 \langle \tilde{b}^{\dagger}_{\vec{k}} \tilde{b}_{\vec{k}} \rangle =
  \int \frac{d^{2} \vec{k} }{ (2 \pi)^{2} } v_{\vec{k}}^{2} $
   which is the quantum depletion of the condensate due to the dipole-dipole
   interaction.

\begin{figure}[tbp]
\includegraphics[width=7cm]{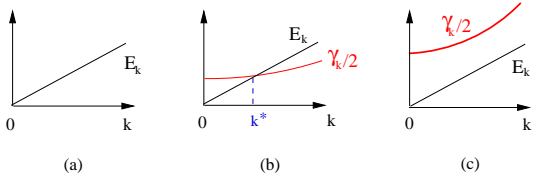}
\caption{The energy spectrum and the decay rate of the exciton
versus in-plane momentum $\vec{k}$.  (a) The energy spectrum of an
equilibrium superfluid with $ \tau_{ex} \rightarrow \infty $. (b)
the indirect exciton superfluid with finite, but large $ \tau_{ex} $
(c) The direct exciton with short $ \tau_{ex} $. } \label{fig6}
\end{figure}

   We can decompose the interaction Hamiltonian $
H_{int}$ in Eqn.\ref{first} into the coupling to the condensate part
$ H_{int}^{c}=\sum_{k_{z}}[ig(k_{z})(\sqrt{N} + \tilde{b}_{0})
a_{k_{z}}+h.c.] $ and to the quasi-particle part $
H_{int}^{q}=\sum_{k}[ig(k)a_{k}\tilde{b}_{\vec{k}}^{\dagger }+h.c.]
$. In the following, we study $ \vec{k}=0 $ and $ \vec{k} \neq 0 $
respectively.

{\bf 3. The photon characteristics at the $\vec{k}=0$ mode.} The
Heisenberg equation of motion of the photon annihilation operator
is:
\begin{equation}
i\partial _{t} \left\langle a_{k_{z}}\right\rangle=(\omega _{k_{z}}-\mu -i\frac{\kappa }{2}%
) \left\langle a_{k_{z}} \right\rangle-ig^{\ast }(k_{z})\sqrt{N}
\label{zero}
\end{equation}%
where the average is over the initial zero photon state $ |in\rangle
= |BEC\rangle |0\rangle_{ph} $ and  $\kappa $ is the decay rate of
the photon due to its coupling to a reservoir  to be determined
self-consistently. The stationary solution of Eqn.\ref{zero} is
\cite{rotation}:
\begin{equation}
\left\langle a_{k_{z}}\right\rangle =\frac{ig^{\ast }(k_{z})\sqrt{N}}{%
(\omega _{k_{z}}-\mu -i\kappa /2)}.
\end{equation}
 which is the photon condensation induced by the exciton
 condensation at $ \vec{k}=0 $ \cite{trivial}. The photon number distribution is
 $ n_{\omega _{k_{z}}}=\left\langle a_{k_{z}}^{\dagger }a_{k_{z}}\right\rangle =%
\frac{N\left\vert g(\mu /c)\right\vert ^{2}}{(\omega _{k_{z}}-\mu
)^{2}+\kappa ^{2}/4} $ where we have set $ g^{\ast }(k_{z})$ around
$\omega _{k_{z}}=\mu $.  The total number of photons is $ N_{ph}=
\sum_{k_z  }n_{\omega_{k_{z}}}= N (|g|^{2}D )/\kappa \label{npht0} $
where $ D= L_{z}/v_{g} $ is the
  photon density of states at $ \vec{k}=0 $.
  Note that the exciton decay rate $ \gamma_0 =|g|^{2}D $ at $ \vec{k}=0 $
  is independent of $ L_{z} $, so is an experimentally measurable quantity.
  The $ N_{ph} $ has to be proportional to $ L_{z} $ in order to get
  a finite photon density in a given volume $ L^{2} \times L_{z} $
  in a stationary state. This self-consistency condition sets $
  \kappa= v_{g}/L_{z} \rightarrow 0 $ so that
\begin{equation}
n_{\omega _{k_{z}}}= N \gamma_{0} \delta( \omega_{k_z}- \mu),
~~~N_{ph}=N \gamma_{0}L_{z}/v_{g} \sim L_{z}  \label{npht}
\end{equation}
  which is independent of $ L_{z} $ as required.
  We showed that the power
  spectrum emitted from the exciton condensate has zero width.
  This conclusion is robust and is
  independent of any macroscopic details such as how photons are
  coupled to reservoirs. We conclude that the coherent light emitted from the condensate has the
  following remarkable properties: (1) highly directional: along the
  normal direction (1) highly monochromatic: pinned at a single
  energy given by the chemical potential $ \mu $ (3) high power:
  proportional to the total number of excitons.
  These remarkable properties could be useful to build highly
  powerful opto-electronic device.

{\bf 4. Input-Output formalism for a stationary state. }
 Now we  discuss the properties of emitted photons
with non-zero in-plane momentum $\vec{k}\neq 0$. It is easy to see
that due to the in-plane momentum conservation, the exciton with a
fixed in-plane momentum $\vec{k}$ coupled to 3 dimensional photons
with the same $\vec{k}$, but with different momenta $k_{z}$ along
the $z$-direction, so we can view these photon acting as the bath of
the exciton by defining $\Gamma _{\vec{k}}=\sum_{k_{z}}g(k)a_{k}$.
When using the standard input-output formalism  which treats $ g(k)
$ non-perturbatively under Markov approximation \cite{book1}, we
find it is convenient to define the input and output fields as:
\begin{eqnarray}
a_{\vec{k}}^{in}(t) &=&\sum_{k_{z}}\frac{1}{\sqrt{D_{\vec{k}}(\omega _{k})}}%
a_{k}(t_{0})e^{-i(\omega _{k}-\mu )(t-t_{0})},  \nonumber \\
a_{\vec{k}}^{out}(t) &=&-\sum_{k_{z}}\frac{1}{\sqrt{D_{\vec{k}}(\omega _{k})}%
}a_{k}(t_{1})e^{-i(\omega _{k}-\mu )(t-t_{1})},  \label{io}
\end{eqnarray}%
where $ t_{0} \rightarrow - \infty $ and $ t_{1} \rightarrow \infty
$ are the initial and final time
respectively, so $ t_0 <t < t_1 $. The density of states of the photon with a given in-plane momentum $%
\vec{k}$ is $ D_{\vec{k}}(\omega _{k})=\frac{\omega
_{k}L}{v_{g}\sqrt{\omega _{k}^{2}-v_{g}^{2}\left\vert
\vec{k}\right\vert ^{2}}}$. It can be shown that the input and
output fields obey the Bose commutation relations
$[a_{\vec{k}}^{in}(t),a_{ \vec{k}^{\prime} }^{in \dagger
}(t^{\prime})]=[a_{ \vec{k}}^{out}(t),a_{ \vec{k}^{\prime}
}^{out\dagger }(t^{\prime})]=\delta_{ \vec{k}, \vec{k}^{\prime} }
\delta (t-t^{\prime})$.

  By solving the Heisenberg equation of motion of Eqn.\ref{first},
  we find that the output field $ a_{\vec{k}}^{out}(\omega )$ is related to the input field by:
\begin{eqnarray}
a_{\vec{k}}^{out}(\omega ) &=&[-1+\gamma _{\vec{k}}G_{n}(\vec{k},\omega +i%
\frac{\gamma _{k}}{2})]a_{\vec{k}}^{in}(\omega )  \nonumber \\
&&+\gamma _{\vec{k}}G_{a}(\vec{k},\omega +i\frac{\gamma _{k}}{2})a_{-\vec{k}%
}^{in\dagger }(-\omega ),  \label{bout}
\end{eqnarray}%
where the normal Green function $ G_{n}(\vec{k},\omega )
=i\frac{\omega +\epsilon _{\vec{k}}+\bar{n}V_{d}( \vec{k})}{\omega
^{2}-E^{2}(\vec{k})} $ and the anomalous Green function
$G_{a}(\vec{k},\omega ) = \frac{i\bar{n}V_{d}(\vec{k})}{\omega
^{2}-E^{2}( \vec{k})} $. The exciton decay rate in the two Green
functions are $\gamma _{\vec{k}}=D_{\vec{k}}(\mu )\left\vert
g_{\vec{k}}(\omega _{k}=\mu )\right\vert ^{2} $ which is independent
of $ L_{z} $, so is an experimentally measurable quantity. Just from
the rotational invariance, we can conclude that $ \gamma _{\vec{k}}
\sim const.+ |\vec{k}|^{2} $ as $ \vec{k} \rightarrow 0 $ as shown
in Fig.1b and 1c. Note that the Fourier transformation of the Eq.
(\ref{io}) leads to $ \omega =\omega _{k}-\mu $. Eqn.\ref{bout} can
be considered as a $ S $ matrix relating the input photon field at $
t_{0} \rightarrow - \infty $ to the output photon field at $ t_{1}
\rightarrow \infty $.

{\bf 5. Photon number spectrum at $ \vec{k} \neq 0 $.} The angle
resolved power spectrum (ARPS) of the output field is $ S_{\pm
}(\vec{k}, \omega )=\int_{-\infty }^{+\infty }d\tau e^{-i\omega \tau
} \left\langle a_{\pm \vec{k}}^{out\dagger }(t+\tau )a_{\pm \vec{k}%
}^{out}(t)\right\rangle _{in} $.  By inserting Eqn.\ref{bout}, one
obtains $S_{\pm }(\vec{k}, \omega )=S_{1}(\vec{k}, \omega )$.
\begin{equation}
S_{1}(\vec{k}, \omega )  =  \frac{ \gamma _{\vec{k}}^{2}
\bar{n}^{2}V_{d}^{2}(\vec{k})}{ \Omega ^{2}(\omega )+\gamma
_{\vec{k}}^{2}E^{2}(\vec{k})} \label{power}
\end{equation}
 where $\Omega (\omega )=\omega^{2}-E^{2}(\vec{k})+\gamma_{\vec{k}}^{2}/4$.
  which is shown in Fig.2a with different $E(\vec{k})$ and $\gamma _{\vec{k}}$.

\begin{figure}
\hspace{-0.4cm}
\includegraphics[width=4.3cm,height=3.4cm]{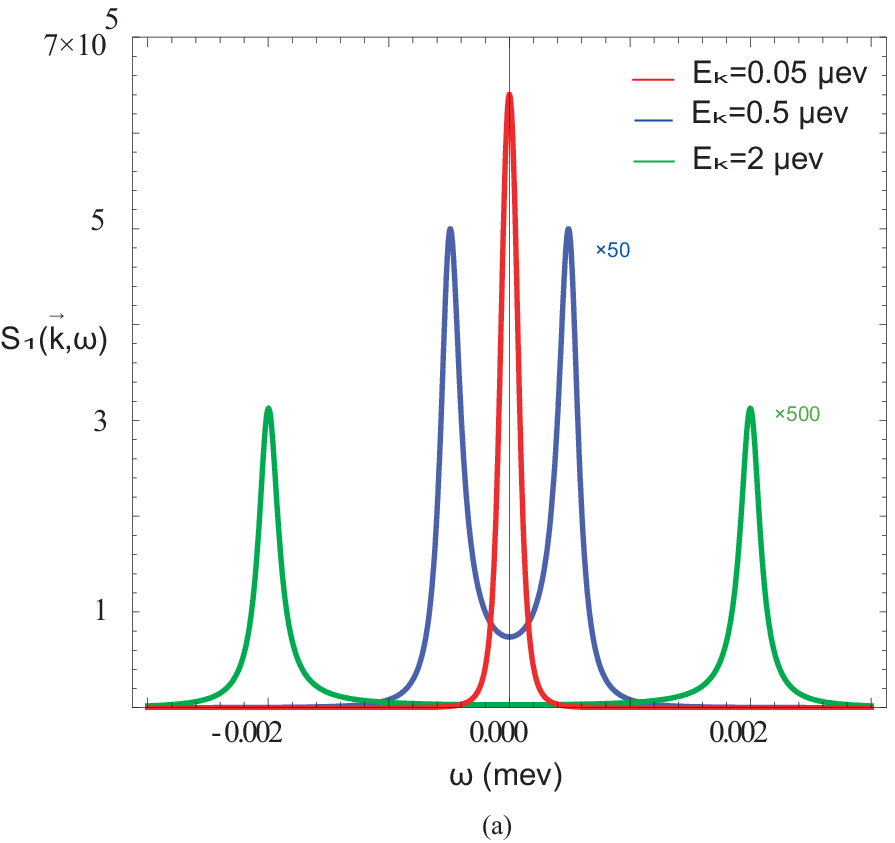}
\hspace{0.1cm}
\includegraphics[width=2cm]{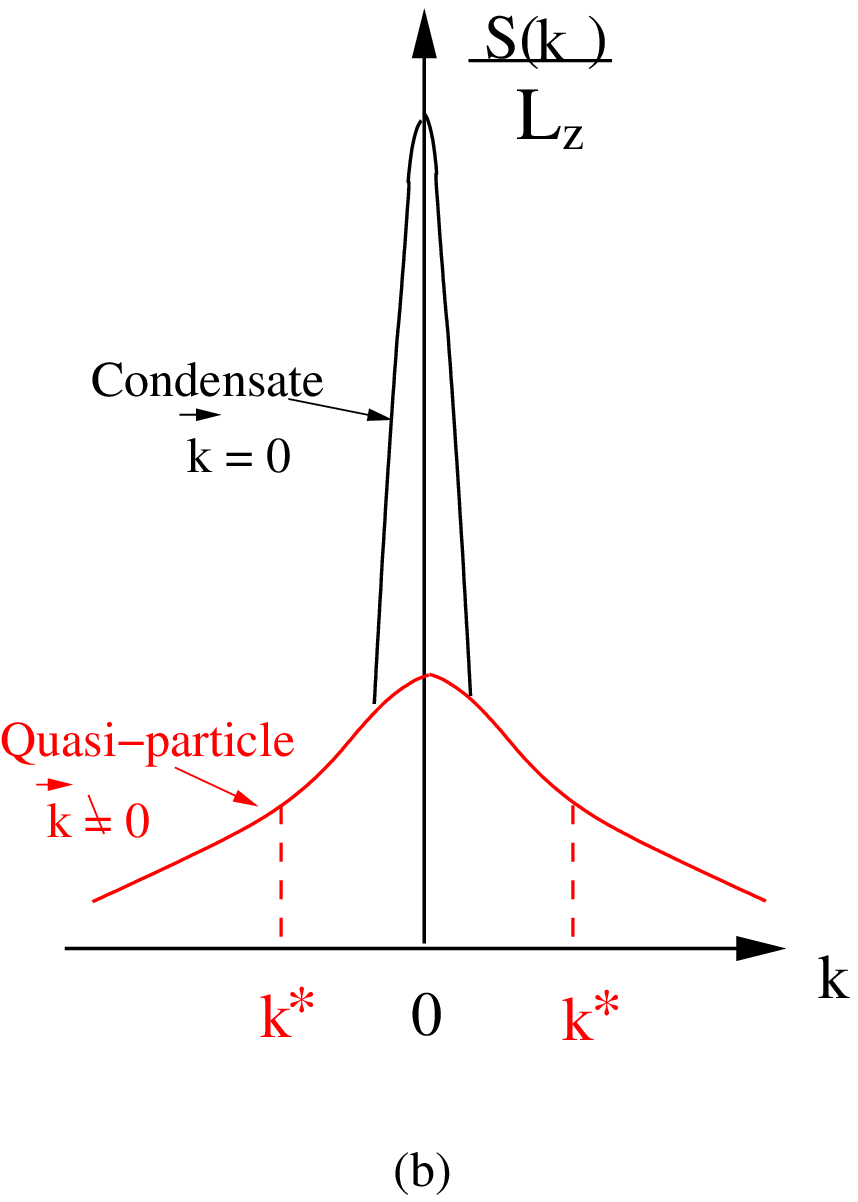}
\hspace{0.1cm}
\includegraphics[width=2cm]{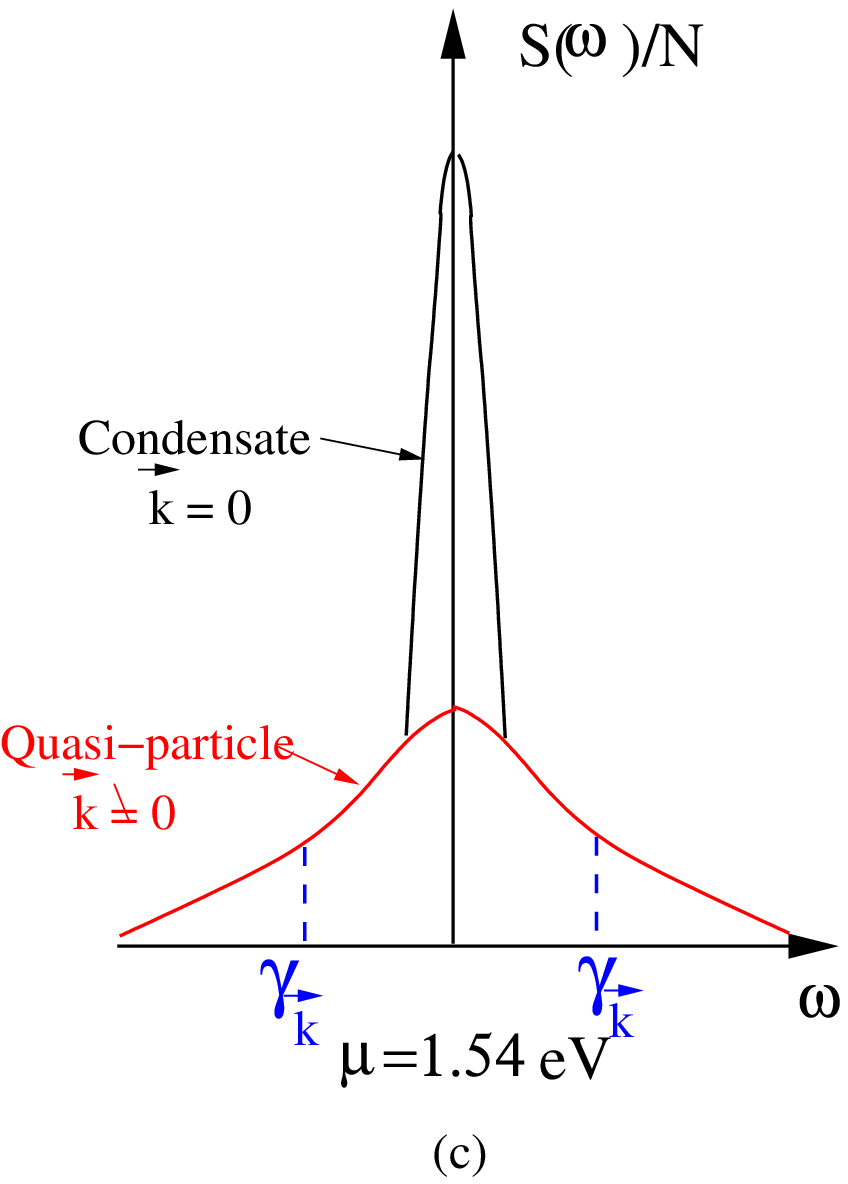}
\caption{ (a) The angle resolved power spectrum (ARPS)  of the
emitted photon with in-plane momentum $ \vec{k} $.  When
$E(\vec{k})\leq \hbar \gamma_{\vec{k}}/2$, there is only one peak
with the width $ \gamma_{\vec{k}}/2 $.  When $E(\vec{k})>\hbar
\gamma_{\vec{k}}/2$, there are two peaks  at the two resonance
photon frequencies $ \omega _{k}=\mu \pm [ E^{2}(\vec{k})-\gamma
_{\vec{k}}^{2}/4]^{1/2} $ also with width $ \gamma_{\vec{k}}/2 $.
The $ S_{1}(\vec{k},\omega ) $ at  $ E( \vec{k})=0.5 \mu eV $ and $
E( \vec{k})=2 \mu eV $ are multiplied by $ 50 $ and $ 500 $ in order
to be seen in the figure. (b) The zero temperature MDC has a
bi-model structure consisting of a sharp peak  $ S( \vec{k}=0 )/L_z=
N \gamma_{0}/v_g $ due to the condensate at $ \vec{k}=0 $
superposing on a Lorentzian peak with a half width $ k^{\ast} \sim
10^{2} cm^{-1} $ due to quasi-particle excitations at $ \vec{k} \neq
0 $. (c) The zero temperature EDC has a bi-model structure
consisting of a sharp $ \delta $ function peak $ S( \omega )/N =
\gamma_{0} \delta( \omega - \mu ) $ due to the condensate at $
\vec{k}=0 $ superposing on a Lorentzian peak with the half envelop
width $ \gamma_{\vec{k}}/2 \sim 0.1 \mu eV $ due to quasi-particle
excitations at $ \vec{k} \neq 0 $.  }
\end{figure}

  In the strong coupling case $ k< k^{*}, E(\vec{k})<\gamma
_{\vec{k}}/2$, $ S_{1}(\vec{k},\omega )$ reaches the maximum $\gamma _{\vec{k}}^{2}\bar{n}%
^{2}V_{d}^{2}(\vec{k})/[\gamma _{\vec{k}}^{2}/4+E^{2}(\vec{k})]^{2}  $ at $%
\omega _{k}=\mu $. As $ \vec{k} \rightarrow 0 $, $
   E(\vec{k}) = u |\vec{k}| \rightarrow 0 $, then $
   S_{1}(\vec{k},\omega ) \rightarrow
   \frac{ \gamma _{\vec{k}}^{2} \bar{n}^{2}V_{d}^{2}(\vec{k})}
   { (\omega ^{2}+ \gamma _{\vec{k}}^{2}/4 )^{2}} $, so the curve has a half
   width $ \sim \hbar \gamma_{0} \sim 10^{-4} meV $. This is
   expected, because the quasiparticles are not well defined
   with the decay rate $ \gamma_{0} $ much larger than its energy $  E( \vec{k})$.
In the weak coupling case $E(\vec{k})>\gamma _{\vec{k}}/2$, at the
two resonance
frequencies $\omega _{k}=\mu \pm \lbrack E^{2}(\vec{k})-\gamma _{\vec{k}%
}^{2}/4]^{1/2}$, $S_{1}(\vec{k}, \omega )$ reaches the maximum $\bar{n}^{2}V_{d}^{2}(%
\vec{k})/E^{2}(\vec{k})$.  It can be shown that when $E(\vec{k}) \gg
\gamma _{\vec{k}}/2$, the width of the two peaks at the two
resonance frequencies is  $ \sim \gamma _{\vec{k}} $, this is
expected, because the quasi-particle is well defined with energy $
\mu \pm [ E^{2}(\vec{k})-\gamma _{\vec{k}}^{2}/4]^{1/2}$ and the
half-width $ \gamma_{\vec{k} }$.

  The Momentum Distribution Curve (MDC)
  $ S_{1}(\vec{k} ) = \sum_{k_z} S_{1}(\vec{k}, \omega ) =
  \int d \omega_{k} D_{\vec{k}}(\omega_k )S_{1}(\vec{k}, \omega )  $ is:
\begin{equation}
   S_{1}(\vec{k} )= \frac{ D_{\vec{k}}(\mu )
   \bar{n}^{2}V_{d}^{2}(\vec{k})\gamma_{\vec{k}}}
   {2[ E^{2}(\vec{k} ) + (\frac{\gamma _{\vec{k}}}{2})^{2}]  }
\label{mdc}
\end{equation}
  which is a Lorentian with the half width
  at $ k^{\ast} \sim 10^{-2} cm^{-1} $ shown in Fig.2b. From Eqn.\ref{npht}, one can see the
  condensate contribution at $ \vec{k}=0 $ is $ N_{ph}/L_{z} = N
  \gamma_{0}/v_{g} \propto N $, while the  contribution from the
  quasi-particle is $ S_{1}(\vec{k} \rightarrow 0 )/L_z = \frac{ 2 n^{2} V^{2}_{d}(0) }{ v_{g} \gamma_{0} }
  \propto n^{2}/\gamma_{0} $. We can also calculate the the one photon
  spacial correlation function:
\begin{equation}
  G_{1}(r) \sim \int \frac{ d^{2} \vec{k} }{ ( 2 \pi )^{2} } \frac{ e^{i
  \vec{k} \cdot \vec{r}} } { k^{2}+ k^{*2}} \sim e^{-k^{*} r }
\label{coh}
\end{equation}
   where we can identify the coherence length $ \xi \sim1/k^{*} \sim
   40  \mu m $. This coherence length has been measured in
   \cite{coherence} and will be discussed in section 7.

 The Energy Distribution Curve (EDC)
 $ S_{1}(\omega ) = \sum_{\vec{k}} S_{1}(\vec{k}, \omega ) $ is:
\begin{eqnarray}
   S_{1}( \omega ) & = &  N \bar{n} \times \int \frac{ d^{2} \vec{k}}{(2 \pi)^{2}}
\frac{ \gamma _{\vec{k}}^{2} V_{d}^{2}(\vec{k})}{ \Omega ^{2}(\omega
)+\gamma _{\vec{k}}^{2}E^{2}(\vec{k})}   \nonumber   \\
    & = & \frac{ N \bar{n} V^{2}_{d}( \vec{k}=0 ) }{ 4 \pi u^{2} } f(
    \frac{ |\omega|}{ \gamma_{\vec{k}=0 } })
\label{edc}
\end{eqnarray}
   where $ f(x)= \frac{1}{x} [ \frac{\pi}{2} -arctg \frac{ 1/4-x^{2}
   }{x} ] $ where $ - \pi/2 < arctg y < \pi/2  $.
   From Eqn.\ref{npht}, one can see the
   condensate contribution at $ \vec{k}=0 $ is $ n_{ \omega_{k_z}}/N =
   \gamma_{0} \delta ( \omega_{k_z} - \mu )  $, while the  contribution from the
   quasi-particle $ S_{1} ( \omega =0 )/N = \frac{ \bar{n} V^{2}_{d}( \vec{k}=0 ) }{  \pi u^{2}
   } $. Both are shown in the Fig.2c.

{\bf 6. Quasi-particle spectrum of a non-equilibrium stationary
exciton superfluid and its experimental observation.  } It is
important to compare the excitation spectra in Fig.1.
  Fig.1a is the well know quasi-particle
  excitations in an equilibrium superfluid. They are well defined
  quasi-particles with infinite lifetime. However, the quasi-particles in
  Fig.1c are not well defined in any length scales, because the decay rate
  is always much larger than the energy. Fig.1b is between the two
  extreme cases. When $ k < k^{*} $, the quasi-particle is not well
  defined, the ARPS is centered around $ \omega_{k}= \mu $ with the
  width $ \gamma_{k} $. The MDC has large values at $ k < k^{*} $.
  The EDC has large values at $ \omega < \gamma_{k} $.
  When $ k > k^{*} $, the quasi-particles is well
  defined,  the ARPS has two well defined quasi-particles peaks at
  $ \omega_{k}=  \mu \pm \lbrack E^{2}(\vec{k})-\gamma _{\vec{k}}^{2}/4]^{1/2}$
  with the width $ \gamma_{k} $. The MDC has very small values at $ k > k^{*} $.
  The EDC has very small values at $ \omega > \gamma_{k} $.
  So in the long wavelength  $ r > \xi \sim 1/k^{*} $ ( or small momentum
  $ k < k^{* } $ ) limit and long time $ \tau > \tau_{ex} \sim 1/\gamma_{k} $
  (or low energy limit $ \omega < \gamma_{k} $ ) limit, there is
  not a well defined superfluid  which is consistent with the
  results achived in \cite{kohn}. However, in the distance  $ r < \xi \sim 1/k^{*} $ ( or momentum
  $ k > k^{* } $ ) limit and the time  $ \tau < \tau_{ex} \sim 1/\gamma_{k} $
  (or  energy scale $ \omega > \gamma_{k} $ ), there is
  still well defined superfluid and associated quasi-particle excitations.
  This is the main difference and analogy between the excitation spectrum in the equilibrium
  superfluid in Fig.1a and that in the non-equilibrium steady state
  superfluid in Fig.1b.  Although we
  derived the Fig.1b from the specific Hamiltonian Eqn.1, we expect
  it is universal for any stationary pumping-decay system such as exciton-polariton systems.
  Very recently, the elementary excitation spectrum of
  exciton-polariton inside a micro-cavity was measured \cite{expb} and
  was found to be very similar to that in a helium 4 superfluid shown
  in Fig.1a except in a small regime near $ k=0 $. We believe this
  observation is precisely due to the excitation spectrum in a
  non-equilibrium stationary superfluid shown in Fig.1b.

{\bf 7. Comments on current experiment data and possible future
experiments.} In \cite{butov,snoke}, the spatially and spectrally
resolved photoluminescence intensity has a sharp peak at the emitted
photon energy $ E=1.545eV $ with a width $ \sim 1 meV  $ at the
lowest temperature $ \sim 1 K \sim 0.1 meV $.  The energy
conservation at $ k_z=0 $ gives the maximum in-plane momentum $
\vec{k}_{max} \sim 1.545eV/v_{g} \sim 4.3\times 10^{4} cm^{-1} $
where we used the speed of the light  $ v_{g}\simeq 8.7\times
10^{9}cm/s$ in $ GaAs $ \cite{butov}. Then the maximum exciton
energy $ E_{max}= u k_{max} \sim 0.15meV $ where we used the spin
wave velocity \cite{butov,snoke} $ u \sim 5 \times 10^{5} cm/s $.
The average lifetime of the indirect excitons in the EHBL
\cite{butov,snoke} is $ \tau_{ex}\sim 40ns $, then we can estimate
the exciton decay rate $ \gamma_{k} \sim \hbar/ 40 ns \sim 10^{-4}
meV =0.1 \mu eV$.  At the boundary of the two regimes in the Fig.1b
where $E(k^{\ast })= u k^{\ast }= \gamma _{k^{\ast }}/2=0.1 \mu eV$,
we can extract $ k^{\ast }=2.4\times 10^{2}cm^{-1} \ll k_{max} $.
The typical  exciton cloud size  $ L \sim 30 \mu m $ \cite{butov}.
The number of excitons is $ N= nL^{2} \sim 10^{5} $ which is
comparable to the number of cold atoms inside a trap in most cold
atom experiments.  The central peak due to the condensate in the MDC
Fig.2b is broadened to $ k_{0} \sim 1/L \sim 10^{3} cm^{-1} $ which
is already larger than the half width due to the quasi-particle $
k^{\ast} \sim 10^{2} cm^{-1} $. So it is impossible to distinguish
the bi-model structure in the MDC in Fig.2b at such a small exciton
size. The coherence length was measured in \cite{coherence}. From
Eqn.\ref{coh}, we find the coherence $ \xi \sim 1/k^{*}\sim 40 \mu m
$ which is slightly larger than the exciton cloud size $ L \sim 30
\mu m $. It is easy to see that the central peak due to the
condensate in the EDC in the Fig.2a is broadened simply due to the
change of the local chemical potential from the center to the edge
of the trap $ \Delta \omega= \frac{1}{2} u_{0} L^{2} \sim 0.1 meV $
where $ u_{0} \sim 10^{-12} eV nm^{-2} $ \cite{butov,snoke}. This
value is much larger than the half width due to the quasi-particle $
 \gamma_{\vec{k}} \sim 0.1 \mu eV$, so it is impossible to distinguish the bi-model structure in
the EDC in Fig.2c either. In order to understand if the observed
peak is indeed due to the exciton condensate, one has to study how
the bi-model structures shown in Fig.2 will change inside a harmonic
trap at a finite temperature $ \sim 1K $ and the effects of both
dark and bright excitons. This will be discussed in a separate
publication. The fine structures in the ARPS in the Fig.2a and phase
sensitive homodyne measurement \cite{un} are also needed to test any
existence of the exciton condensate.

{\bf 8. Conclusions } We study the power spectrum  of photons
emitted from the exciton superfluid phase in semiconductor
electron-hole bilayer systems. We find that the  photons emitted
along the direction perpendicular to the layer are in a coherent
state which possesses several remarkable properties, while those
along all tilted directions due to the quasi-particles above the
condensate show very interesting structures. We determined the angle
resolved power spectrum ( ARPS ), the line shapes of both the MDC
and the EDC. We also pointed out the analogy and difference between
the quasi-particles in an equilibrium superfluid and those in a
non-equilibrium stationary superfluid. This difference precisely
explained the recent experimental observation of excitation spectrum
of exciton polariton in a planar microcavity. We commented on
available experimental data both MDC and EDC and also suggested
possible future ARPS experiment to test our theoretical predictions

We are very grateful for C.P. Sun for many helpful suggestions and
encouragements. We also thank Peng Zhang for helpful discussions.
J.Ye is indebted to B. Halperin for critical reading of the
manuscript and many helpful suggestions. J. Ye is grateful for A. V.
Balatsky, L. Butov, Jason Ho, Guoxinag Huang, Xuedong Hu, Allan
Macdonald, G. Murthy, Zhibing Li, Qian Niu, Zhe-Yu Oh, Lu Sham, D.
Snoke, Hailing Wang, Congjun Wu, C. L Yang, Wang Yao, Fuchun Zhang,
Weiping Zhang for helpful discussions. J. Ye's research at KITP-C
was supported by the Project of Knowledge Innovation Program (PKIP)
of Chinese Academy of Sciences; at KITP was supported in part by the
NSF under grant No. PHY-0551164.



\begin{thebibliography}{99}

\bibitem{blatt} John M. Blatt, K. W. B\"{o}er and Werner Brandt, Phys. Rev.
126, 1691 (1962).

\bibitem{kel} L. V. Keldysh and A. N. Kozlov.
\textsl{Sov, Phys. JETP} 27, 521-528 (1968).

\bibitem{loz} Yu. E. Lozovik and V. I. Yudson, Pis'ma Zh. Eksp. Teor. Fiz. 22, 556
(1975) [JETP Lett. 22, 274 (1975)]

\bibitem{butov} L. V. Butov, {\sl et al},
Nature 417, 47 - 52 (02 May 2002); Nature 418, 751 - 754 (15 Aug
2002); C. W. Lai, {\sl et al} Science 23 January 2004 303: 503-506.


\bibitem{snoke} D. Snoke, {\sl et al }  Nature 418, 754
- 757 (15 Aug 2002); David Snoke, Nature 443, 403 - 404 (28 Sep
2006);

\bibitem{field1} U. Sivan, P. M. Solomon, and H. Shtrikman,  Phys. Rev. Lett. 68, 1196 - 1199 (1992)

\bibitem{field2} J. A. Seamons, D. R. Tibbetts, J. L. Reno, M. P. Lilly,
arXiv:cond-mat/0611220.

\bibitem{bell} R. Rapaport, {\sl et al}, Phys. Rev. Lett. 92, 117405 (2004);  Phys. Rev. B 72, 075428
(2005).

\bibitem{coherence} Sen Yang, A. T. Hammack, M. M. Fogler, and L. V. Butov,
Phys. Rev. Lett 97, 187402 (2006).

\bibitem{expb}  S. Utsunomiya, {\sl et al },
 Nature Physics 4, 700 - 705 (01 Sep 2008).

\bibitem{un} Tao Shi, Jinwu Ye, Longhua Jiang and C.P. Sun,
unpublished.


\bibitem{ye} Jinwu Ye,  arXiv:cond-mat/0712.0437.


\bibitem{psdw} Jinwu Ye and Longhua Jiang, Phys. Rev. Lett. 98, 236802
(2007), Jinwu Ye, Phys. Rev. Lett. 97, 236803 (2006),  Jinwu Ye,
Annals of Physics, 323 (2008), 580-630.







\bibitem{andrei} P. Mehta and N. Andrei,
 Phys. Rev. Lett. 100, 086804 (2008); {\sl ibid}, 96, 216802 (2006).



\bibitem{rotation} In order to keep the relative energy difference
between the exciton and the photon intact, we made a rotation
 $a=  \tilde{a}
e^{-i \mu t} $ and neglected the $ \tilde{} $ in the following.


\bibitem{trivial} This straightforward conclusion was reached in
Y. Yamamoto and A. Imamoglu, Mesoscopic quantum optics, John Wiley
\& Sons, Inc. 1999. P213. A. Castro {\sl et.al}, PRL 87, 246403. But
the  results achieved at $ \vec{k}=0 $ in the rest of the following
section are highly non-trvial and completely new. The results on $
\vec{k} \neq 0 $ achieved in the other sections are also completely
new.

\bibitem{book1} D. F. Walls and G. J. Milburn, Quantum Optics,
Springer-Verlag, 1994. Chap.7.


\bibitem{kohn} W. Kohn and D. Shrrington, Rev. Mod. Phys. 42, 1-11
(1970).






\end{thebibliography}
\end{document}